%% file: paper.tex
\documentclass[conference]{IEEEtran}
\IEEEoverridecommandlockouts
\usepackage[utf8]{inputenc}
\usepackage[T1]{fontenc}
\usepackage{microtype}
\usepackage{listings}
\usepackage{makecell}
\usepackage{xargs}
\usepackage{lipsum}
\usepackage{xparse}
\usepackage{xifthen, xstring}
\usepackage{xspace}
\usepackage{marginnote}
\usepackage{etoolbox}
\usepackage[acronym,shortcuts]{glossaries}
\usepackage{amsmath}
\usepackage{thmtools} 
\usepackage{algorithm}
\usepackage[noend]{algpseudocode}
\usepackage{hyphenat}
\usepackage[shortcuts]{extdash}
\usepackage{url}

\input{setup.tex}

\makeatletter
\ifcsdef{MintedExecutable}
{
  \newminted[mlir]{tools/lexers/MLIRLexer.py:MLIRLexerOnlyOps}{mathescape}
  \newminted[xdsl]{tools/lexers/MLIRLexer.py:MLIRLexer}{mathescape, style=murphy}
  \newminted[lean4]{tools/lexers/Lean4Lexer.py:Lean4Lexer}{mathescape}
  \newmintinline[mlirinline]{tools/lexers/MLIRLexer.py:MLIRLexerOnlyOps}{mathescape}
  \newmintinline[xdslinline]{tools/lexers/MLIRLexer.py:MLIRLexer}{mathescape, style=murphy}
}
{
    \ifcsdef{minted@optlistcl@quote}
  {
    \newminted[mlir]{tools/lexers/MLIRLexer.py:MLIRLexerOnlyOps}{customlexer, mathescape}
    \newminted[xdsl]{tools/lexers/MLIRLexer.py:MLIRLexer}{customlexer, mathescape, style=murphy}
    \newminted[lean4]{tools/lexers/Lean4Lexer.py:Lean4Lexer}{customlexer, mathescape}
    \newmintinline[mlirinline]{tools/lexers/MLIRLexer.py:MLIRLexerOnlyOps}{customlexer, mathescape}
    \newmintinline[xdslinline]{tools/lexers/MLIRLexer.py:MLIRLexer}{customlexer, mathescape, style=murphy}
  }
  {
    \newminted[mlir]{tools/lexers/MLIRLexer.py:MLIRLexerOnlyOps -x}{mathescape}
    \newminted[xdsl]{tools/lexers/MLIRLexer.py:MLIRLexer -x}{mathescape, style=murphy}
    \newminted[lean4]{tools/lexers/Lean4Lexer.py:Lean4Lexer -x}{mathescape}
    \newmintinline[mlirinline]{tools/lexers/MLIRLexer.py:MLIRLexerOnlyOps -x}{mathescape}
    \newmintinline[xdslinline]{tools/lexers/MLIRLexer.py:MLIRLexer -x}{mathescape, style=murphy}
  }
  
}
\makeatother

\begin{document}

\title{Lifting to tensors when compiling scientific computing workloads for AI Engines}

\author{\IEEEauthorblockN{Nick Brown}
\IEEEauthorblockA{\textit{EPCC at the University of Edinburgh} \\
Edinburgh, UK \\
n.brown@epcc.ed.ac.uk}
\and
\IEEEauthorblockN{Gabriel Rodriguez-Canal}
\IEEEauthorblockA{\textit{EPCC at the University of Edinburgh} \\
Edinburgh, UK}
}

\maketitle

\begin{abstract}
It has been demonstrated that specialised architectures, such as FPGAs and AMD's AI Engines (AIEs), have the potential to deliver energy and performance advantages for scientific computing. Given the integration of AIEs into AMD's CPUs, this is an interesting potential avenue especially when executing on the edge or making better use of local compute constrained resources. However, a major challenge is in enabling existing codes to run on this architecture without extensive modification. Put simply, it requires significant expertise and time to port codes to the AIE's execution model.
  
In this paper we explore a compilation pipeline for efficiently mapping loops in general purpose, scientific codes to AIEs. Lifting the semantics of an application into tensors, we demonstrate that this is able to capture the intention of general purpose loops annotated with OpenMP and such high-level tensor information provides a richness that is effective when mapping to the AIEs. Requiring only an OpenMP decorated loop, our approach significantly reduces code complexity when targeting the architecture. For six kernel benchmarks, representing AI and scientific computing, using our approach the NPU performs comparatively to the multicore CPU for float32, in all cases at reduced energy to solution. For two scientific computing kernels running across both the CPU and NPU together delivers up to a 40\% improvement in performance and 15\% reduction in energy usage compared to the CPU alone.
\end{abstract}

\begin{IEEEkeywords}
MLIR, AMD AI Engines, HPC, tensors
\end{IEEEkeywords}

\section{Introduction}
\label{sec:intro}
High Performance Computing, HPC, is heavily used for scientific computing workloads. Whilst many people might naturally assume leadership class supercomputers, more constrained local clusters are also popular for scientific computing workloads and the edge is also worthwhile exploring. HPC relies on mainstream CPUs and GPUs, but with an emphasis on energy efficiency and the continuing importance of delivering increasing performance to meet the ever growing demand from users, there is interest in leveraging specialised hardware technologies typically designed for AI/ML. AMD Xilinx's AI engines are one example, initially released as part of the Versal Adaptive SoC these are vector arithmetic accelerators. AI Engines, or AIEs, adopt a Very Long Instruction Word (VLIW) design and contain a dedicated 512 bit vector unit. It has been demonstrated that there is potential for these in HPC \cite{brown2023exploring} \cite{klaisoongnoen2024evaluating}, and in 2023 AMD released the Ryzen AI series of CPUs which combines AIEs, termed the Neural Processing Unit (NPU), with traditional x86 cores. 

Since then AMD have released a range of NPUs, based on the XDNA v1 and v2 architectures. The close coupling between the NPU and CPU cores offers a range of potential opportunities for leveraging this specialised compute, intended primarily for AI/ML, to accelerate scientific computing. However, a major challenge is in the programming of this architecture. Whilst AMD have made progress via IRON \cite{hunhoff2025efficiency}, one must still rewrite codes and recast algorithms into a form that is suitable for the AIE. This is not only time consuming, but also requires extensive expertise. Furthermore, existing AIE programming approaches are Python and C++ based which, in a world where around 60\% of scientific computing codes are written in Fortran \cite{rodriguez2023fortran}, is a challenge in and of itself.

In this paper we describe an approach which enables general purpose loops to be seamlessly offloaded to the NPU. Focussing on Fortran and OpenMP, two very popular programming technologies in the scientific computing community, by lifting the representation of a loop to tensors our approach is able to exploit this high-level information when making decisions around how to efficiently target the hardware. 



The contributions of this paper are as follows:
\begin{itemize}
  \item We demonstrate much of the MLIR tensor representation is applicable beyond AI/ML, and transformations are able to lift OpenMP loops to this representation.
  \item We highlight how, based upon the rich information present in the tensor representation, the compiler can make effective decisions when mapping general purpose, loop based, computing workloads to the NPU. 
  \item We demonstrate that the tight coupling on Ryzen-AI between CPU and NPU delivers performance and energy benefits when co-executing loop iterations.
\end{itemize}

\section{Background and related work}
\label{sec:bg}

\begin{figure}[htb]
\centering
 \includegraphics[width=0.8\columnwidth]{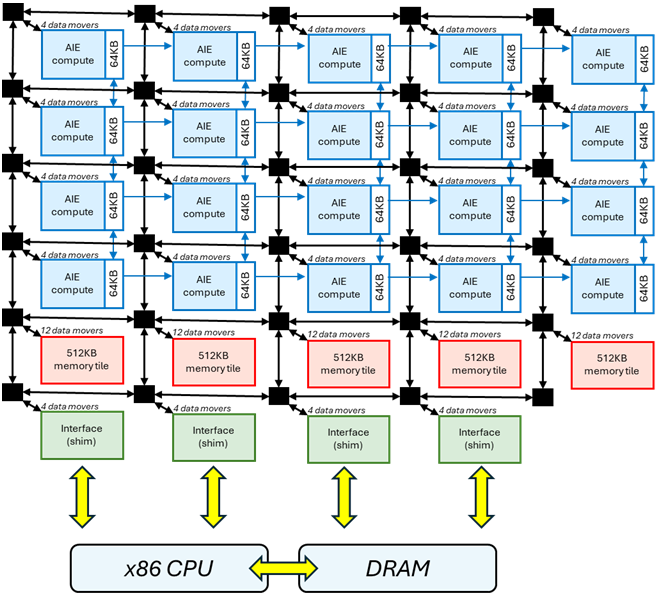}
\caption{Illustration of AMD's Hawk Point NPU, comprising five columns of four rows of AIEs (each with compute core and 64KB of memory). Each column has a 512KB memory tile and four columns have an interface tile.}	
\label{fig:hawkpoint}
\end{figure}

AMD Embedded, formerly Xilinx, introduced the AI Engine (AIE) to provide hardened support for common arithmetic operations on their FPGAs. Each AIE employs a very-long-instruction-word (VLIW) architecture capable of issuing up to seven instructions per cycle, and supports both scalar and vector execution via a 512-bit vector unit. In 2023 AMD Launched their XDNA Phoenix Neural Processing Unit (NPU) which integrates the AIEs directly into the Ryzen AI CPU. Figure \ref{fig:hawkpoint} sketches the XDNA v1 architecture, which we focus on in this paper, where NPU tiles are organised as a two-dimensional mesh with nearest-neighbour connectivity in both dimensions. An AIE can directly access the local memories of its north, south, and west neighbours. Each engine also incorporates four data movers comprising two 32-bit input streams and two 32-bit output streams.

As sketched in Figure \ref{fig:hawkpoint}, the Hawk Point NPU we focus on in this paper comprises 20 AIEs arranged as five columns by four rows. Each column additionally incorporates a memory tile, and four of the columns include an interface (shim) tile connecting the array to the CPU and main memory. The NPU uses the AIE‑ML generation of AI engine which has been optimised for AI workloads and provides a per‑engine DRAM memory of 64 KB. Each dedicated memory tile contains 512 KB of SRAM and 12 data movers, in contrast to the four provided by compute and interface tiles, delivering up to 30 GB/s of aggregate bandwidth \cite{aie-ml}. XDNA vector units do not natively support int32 or float32 datatypes which must be emulated. This is a limitation but AI engines are evolving rapidly and coupling the x86 CPU cores with NPU is promising. 

In our opinion, AIEs integrated with an existing x86 CPU are an attractive proposition for general purpose programmers as they can still leverage the CPU for their code as normal, and the tight integration with the NPU delivers the possibility of accelerating key parts of the code. Software written for the AIEs comprises two parts, the compute kernels which are mapped to the AI Engines and a graph description that connects interfaces, kernels and memories together via streams. Kernels follow a producer–consumer model, consuming input from up to two streams and emitting results on up to two output streams. Streams can connect directly to the CPU through an interface tile, to a memory tile, or to another compute tile.

There have been a variety of successes in using AMD's AIEs for HPC workloads, for instance \cite{bouaziz2025dataflow} demonstrated significant performance benefits when running option pricing on AIEs compared to CPUs. Furthermore, a fundamental operator of CNNs was accelerated using the AIEs on the Versal ACAP in \cite{zhang2023new}, providing up to a 139 times speed up compared to the CPU. Other efforts have looked to address programmer productivity on the AIEs, for instance \cite{levental2024end} proposed an end to end AIE programming model that leverages a Python Domain Specific Language (DSL) and the HPX programming framework was enhanced to support AIEs in \cite{kalkhof2024enabling}. However, both these require rewriting codes into their respective frameworks. A subset of Fortran intrinsic subroutines were offloaded to AIEs in \cite{brown2025seamless} and whilst that work only supports a very restrictive set of workloads that heavily use Fortran intrinsics, it demonstrates the composability benefits of MLIR.

\subsection{LLVM and MLIR}

LLVM is a modular, reusable compiler and toolchain infrastructure that enables the construction of compilers for diverse programming languages and hardware targets. It provides language front-ends and a broad set of hardware back-ends, connected via the LLVM intermediate representation (LLVM IR). A front-end, such as Flang, emits LLVM IR and can, in principle, target any supported back-end, including CPUs, GPUs, and FPGAs. However, LLVM IR is low level and substantial effort is required, and duplication common, in each front-end when generating LLVM IR.

To this end, Google developed MLIR and released it open source in 2019. MLIR provides a multi-dialect intermediate representation with standard transformations between these dialects. Instead of lowering directly to LLVM IR, front-ends translate into one or more higher-level dialect representations and then rely on existing MLIR passes to progressively lower to LLVM IR. MLIR uses the standard Static Single Assignment (SSA) form for IR, and a key advantage is that dialects can be composed and manipulated independently, enabling staged lowering that incrementally moves towards the target architecture. MLIR promotes extensive sharing of compiler infrastructure by reusing established dialects and transformations, substantially reducing development effort. Core dialects include \emph{memref} for memory, \emph{func} for functions and \emph{arith} for arithmetic, and furthermore MLIR provides a framework for defining custom dialects and transformations. Indeed, AMD have added several MLIR dialects to support compilation for the AIE such as \emph{aie} which describes streaming connections between AIE compute tiles and Direct Memory Access (DMA) and \emph{adf} to express AMD's Adaptive Data Flow (ADF) graph connecting tiles. Moreover, a range of transformations and optimisations have also been developed by AMD, enabling lowering from these dialects to instructions that will execute on the NPU.

\begin{figure*}[htb]
\centering
 \includegraphics[width=\textwidth]{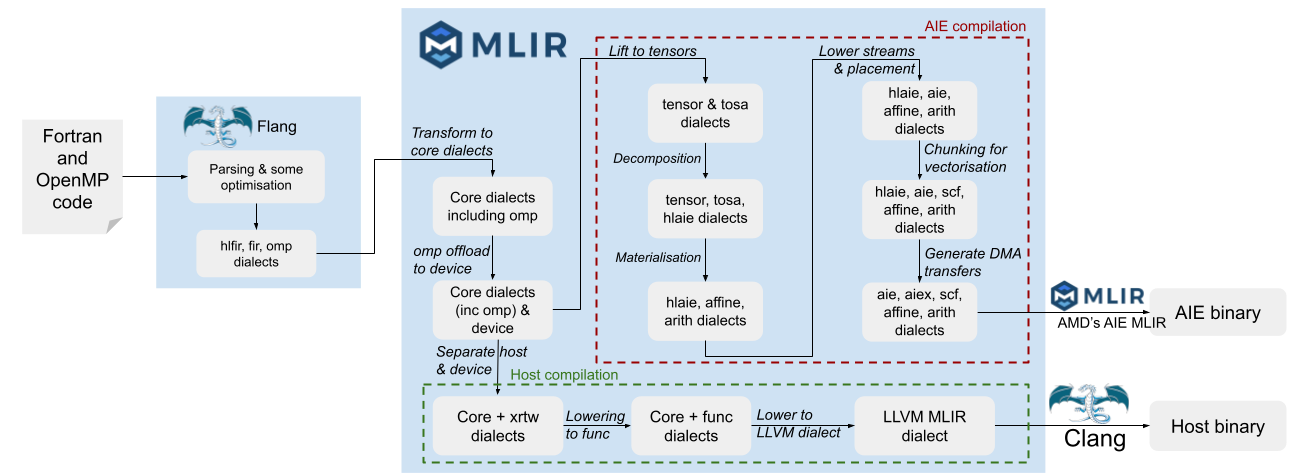}
\caption{Illustration of our MLIR-based OpenMP loop compilation flow for the AI Engines.}	
\label{fig:our-flow}
\end{figure*}

Flang is the LLVM project's Fortran front-end and built from the ground up using MLIR. Aiming to provide comprehensive support for the Fortran language, Flang generates IR based upon the \emph{hlfir} (High Level Fortran Intermediate Representation) and \emph{fir} (Fortran Intermediate Representation) dialects. It then undertakes a series of transformation and optimisation passes on these dialects before generating LLVM IR. Flang sits outside of the core MLIR ecosystem and integrates with only a subset of the core dialects, where \emph{hlfir} \& \emph{fir} are not part of MLIR itself, instead with Flang providing its own path to LLVM IR. Consequently, \cite{brown2024fully} developed a lowering from \emph{hlfir} \& \emph{fir} into core MLIR dialects. It was found that integrating with the entire MLIR ecosystem provides some performance benefits, but also crucially flexibility. For example a much wider range of dialects are available, including those provided by vendors for architectures such as the NPU. 

\subsection{OpenMP}

Since standardisation in 1997, OpenMP has become the de-facto model for shared-memory parallelism in scientific computing. Originally designed for threaded CPU workloads, 2017 saw the introduction of the \emph{target offload} directive as part of OpenMP 4.0 to suppirt accelerators, primarily GPUs. Driven by Fortran and C, Listing \ref{lst:ftn-eg} sketches an example Fortran loop decorated with OpenMP \emph{target} pragma for offload.

MLIR provides support for OpenMP via the \emph{omp} dialect. There have been several efforts to target FPGAs with OpenMP, for instance Nymble \cite{huthmann2020openmp} and \cite{openmpfpga}. By comparison, instead of FPGAs we target the AI engines that are already present in Ryzen AI CPUs. The cornerstone of our approach is to lift the abstraction level from OpenMP to the MLIR \emph{tensor} dialect, and our hypothesis is that this will then provide a rich source of information for the compiler. Tensorize \cite{brauckmann2025tensorize} is a compiler approach lifting legacy code to a tensor representation, in their case the linear algebra \emph{linalg} MLIR dialect. From this it generates NumPy or StableHLO. Our approach differs in two main ways, firstly \cite{brauckmann2025tensorize} only targets CPUs and GPUs whereas our focus is on other architectures, such as the NPU, which are very different and whose compiler support is less mature for general purpose programming. Secondly, Tensorize comprises a significant amount of complexity in mapping a range of loops in Python or C. Instead, by requiring the programmer to have decorated their loops with OpenMP then our lifting to tensors is significantly simplified because we are able to leverage guarantees, such as the independence of loop iterations, provided by the corresponding OpenMP pragmas. 

\section{A loop based OpenMP AIE compiler flow}
\label{sec:accelerate}

Figure \ref{fig:our-flow} illustrates our compiler flow where Fortran code, annotated with OpenMP target offload, is first processed by Flang which generates IR that comprises mainly the \emph{hlfir}, \emph{fir} and \emph{omp} dialects. As described in Section \ref{sec:bg}, \emph{hlfir} and \emph{fir} are then transformed by \cite{brown2024fully} to core MLIR dialects including \emph{omp}. Whilst we use Fortran as a driver, our work is not tied to that front-end and-so this provides future flexibility. 

The contribution of our work starts at \emph{omp offload to device} in Figure \ref{fig:our-flow}, where operations in the \emph{omp} dialect are transformed into our \emph{device} dialect which simplifies the mapping to host side XRT calls. Data transfers between the host and device are transformed into \mlirinline{device.alloc} which returns \emph{memrefs} that are tagged with an explicit memory space on the device. Host and device IR is then separated into separate modules, and host-side \emph{device} dialect operations are then lowered into the XRT Wrapper MLIR dialect, \emph{xrtw}, from \cite{sc25-intrinsic}. Ultimately, this results in LLVM IR on the host comprising the non-accelerated Fortran code and calls to XRT for interacting with the NPU, which is then compiled by Clang to generate the host binary.

The module comprising code to run on the AIEs contains OpenMP operations that describe the structure of the loops, for example \mlirinline{omp.parallel} for a parallel loop and \mlirinline{omp.loop_nest} is a nested loop. Modifiers, such as \mlirinline{omp.private} for thread private data and \mlirinline{omp.declare_reduction} are also present in the IR. This information is now leveraged by the transformation \emph{lift to tensors} in Figure \ref{fig:our-flow} that lifts the IR into the \emph{tensor} and \emph{tosa} (Tensor Operator Set Architecture) dialects. The \emph{tensor} dialect handles generic creation and manipulation of tensors, whereas the \emph{tosa} dialect implements the TOSA specification \cite{tosa} which provides a set of common machine learning operations.

Listing \ref{lst:ftn-eg} sketches a simple Fortran loop using OpenMP \emph{target} offload which adds each element in the \emph{a} and \emph{b} arrays before multiplying the result with a constant stored in \emph{c}. Listing \ref{lst:lifted-ir} portrays the IR in the \emph{tensor} and \emph{tosa} dialects that has been lifted from this Fortran code. Tensors provide value semantics, where the focus is on the values rather than the concrete implementation and the \mlirinline{device.tensor_compute} operation wraps all tensor operations in the IR. This provides a bridge between abstraction levels from reference semantics elsewhere to the high-level value semantics of tensors in this lifted form.

\begin{listing}[h]
\begin{lstlisting}[language=Fortran, frame=none, numbers=left]
!$omp target parallel do private(t) map(from:c)
do i=1, 128
  t=a(i)+b(i)
  c(i)=t*100
end do
!$omp end target parallel do
\end{lstlisting}
\caption{Example Fortran loop offloaded with OpenMP
\label{lst:ftn-eg}}
\end{listing}

The \mlirinline{tosa.add} and \mlirinline{tosa.mul} operations in Listing \ref{lst:lifted-ir} perform element wise addition and multiplication respectively, with \mlirinline{tensor.splat} broadcasting a scalar value into each element of a tensor. The resulting tensor from this computation, \emph{c}, is then yielded as the result of the \mlirinline{device.tensor_compute} operation. Tensors and TOSA provides a rich representation around the intention of compute without complications at this stage of lower details such as to how it will be achieved.

\begin{listing}[htb]
\begin{mlir*}{}
^0(
                      tensor<128xf32>
                      tensor<128xf32>
  device.tensor_yield(
}) : (memref<128xf32>, memref<128xf32>) 
         -> tensor<128xf32>
\end{mlir*}
\caption{Sketch of IR based on Fortran loop in Listing \ref{lst:ftn-eg} lifted to the tensor and tosa dialects. \label{lst:lifted-ir}}
\end{listing}

Tensors are capable of capturing a wide range of computation and data access constructs. For example, Listing \ref{lst:stencil-ir} sketches how stencil calculations such as \emph{c[i] = a[i-1] + b[i+1]} are represented in tensors, where \emph{a\_e} and \emph{b\_e} are extracted from the \emph{a} and {b} tensors respectively using \mlirinline{tensor.extract_slice}. The indexes \emph{[0] [128] [1]}, for instance in the first \mlirinline{tensor.extract_slice} operation, denote the offset, number of elements and stride. A \mlirinline{tosa.add} operation then performs an element wise addition of these slices, with \emph{a\_e} containing 128 elements from index 0, and \emph{b\_e} 128 elements starting from index 2. The resulting slice \emph{res} is then inserted into \emph{c} by the \mlirinline{tensor.insert_slice} operation which produces the final resulting \emph{c\_res} tensor.

\begin{listing}[htb]
\begin{mlir*}{}
       : tensor<130xf32> to tensor<128xf32>
       : tensor<130xf32> to tensor<128xf32>
        into tensor<130xf32>
\end{mlir*}
\caption{Sketch of tensor based IR for calculating \emph{c[i] = a[i-1] + b[i+1]} \label{lst:stencil-ir}}
\end{listing}

OpenMP imposes certain restrictions on loops, such as loop iteration independence, and this makes lifting to tensors more straight forwards than in Tensorize \cite{brauckmann2025tensorize} which lifted legacy code to a tensor representation in the \emph{linalg} dialect. At the implementation level, our transformation pass identifies the outputs of the loop and, for each of these, walks the IR backwards to build up a dependency graph of operations connecting loop inputs to outputs. A conversion is then undertaken for each constituent operation within each graph to generate its tensor counterpart. Whilst this handles a wide range of loop structures, we do not currently support atomic OpenMP pragams and the presence of these will cause the loop to fallback to the CPU. 

Listings \ref{lst:lifted-ir} and \ref{lst:stencil-ir} illustrate how the tensor representation provides a rich description of the compute within loops. This can then be used to drive decisions when targetting the AIEs. For instance, the offsets in Listing \ref{lst:stencil-ir} influence how FIFOs are generated and tensors enable dependencies between operations to be discovered. The next transformation in the pipeline of Figure \ref{fig:our-flow}, \emph{decomposition}, uses this dependency information to determine placement of compute across the NPU. We provide two strategies; decomposing operations and/or decomposing loop iterations across the NPU. Mixing of these strategies is supported, for instance in Listing \ref{lst:lifted-ir}, the \mlirinline{tosa.mul} operation might be placed on one AIE and \mlirinline{tosa.add} on another, and these groups of two AIEs replicated across four, each acting on a unique chunk of iterations. Limitations imposed by the architecture restrict and influence these decisions, most importantly that compute tiles have a maximum of two inputs and two outputs. Crucially in our approach the compiler handles this rather than it being the programmer's responsibility.

At this stage tensor operations and/or loop iterations have been distributed across the NPU and this is represented by our high-level dialect, \emph{hlaie}. The \emph{hlaie} dialect is a step down in abstraction from tensors, and encodes the decomposition across the NPU and AIE interactions, but not how these are achieved. The dialect comprises the following operations:

\begin{enumerate}
    \item \mlirinline{hlaie.kernel} defines a compute kernel, taking up to two \emph{hlaie.stream}s as input and up to two as results.
    \item \mlirinline{hlaie.memory} represents a memory tile.
    \item \mlirinline{hlaie.external} host and device connection.
    \item \mlirinline{hlaie.stream}s values between tiles.
    \item \mlirinline{hlaie.stream_read} reads value(s) from a stream.
    \item \mlirinline{hlaie.stream_write} writes value(s) to a stream.
\end{enumerate}

At this point in the compilation pipeline the IR contains \mlirinline{hlaie.kernel}, \mlirinline{hlaie.memory} and \mlirinline{hlaie.external} operations. Compute has been distributed across the NPU and each of these contains specific tensor operations, with tile level inputs and outputs connected via \mlirinline{hlaie.stream}. The \emph{materialisation} pass of Figure \ref{fig:our-flow} lowers from value semantics of tensors into reference semantics of \emph{affine} loops that read specific values from stream(s) and the \emph{arith} dialect then performs arithmetic upon these, with results then written via \mlirinline{hlaie.stream_write} to output stream(s). 

The next transformation pass, \emph{lower streams \& placement} in Figure \ref{fig:our-flow}, materialises the kernel, memory and external operations to actual AIE tiles. This involves mapping to physical compute, memory and shim tiles and making decisions around placement. We aim to place components that communicate on tiles near each other, for instance mapping \mlirinline{hlaie.kernel}s that stream data to neighbouring \mlirinline{aie.core}s. After exploiting the dependency information held by the \emph{hlaie} dialect these operations are then lowered into corresponding FIFO operations within AMD's \emph{aie} dialect. 

Our approach then executes the \emph{chunking for vectorisation} transformation which manipulates the inner structure of each kernel following \cite{aie-vect} to vectorise arithmetic operations. It inserts an inner \mlirinline{affine.for} loop of iteration count vector width, and an outer loop stepping from one chunk to the next. Lastly, DMA transfer operations are generated which determines DMA between the host and device, both driving the streams and copying data into memory tiles. A complication was that, as part of this work, we discovered AMD's MLIR flow itself does not support vectorisation and this impacts performance on the NPU as AIEs run scalar only. Consequently, after \emph{chunking for vectorisation}, transformed IR is extracted and provided to AMD's \emph{aie-translate} tool which generates vectorised C++ code using AIE intrinsics from the IR. The rest of the IR is provided to AMD's \emph{aie-opt} tool which performs a series of further passes to lower from the AMD specific \emph{aie}, \emph{aievec} and \emph{aiex} dialects to LLVM IR. The generated C++ and remaining LLVM IR is then compiled with either AMD's open source Peano compiler or Chess, our approach providing automatic vectorisation of user code. 

The key point of this section is that lifting to tensors is the key enabler here and the tensor abstraction is able to capture the compute pattern of loops, especially when driven by OpenMP due to the guarantees provided by the semantics of those operations. Other alternatives, such as mapping loops to the \emph{affine} dialect, lack the Destination Passing Style (DPS) of \emph{tensor} and \emph{tosa} dialects which provide a richer description of data and dependencies when mapping to the AIEs. Ultimately, this rich information can then be exploited by the compiler to make sensible decisions around how to effectively target the NPU. By lowering the abstraction levels from tensors, through our \emph{hlaie} dialect, ultimately to AMD's AIE dialects we are able to progressively materialise the key aspects in the IR to suit the architecture.

\section{Results and evaluation}
\label{sec:performance}
Our experiments run on an eight core Ryzen AI 8945HS CPU equipped with 64GB of DRAM and containing the Hawk Point, XDNA v1, NPU. We use GCC version 14, XRT release version 2025.1, Flang, LLVM and MLIR versions 20.1.7, and the release version 1.1.0 of AIE-MLIR. All results are averaged over ten runs and AIE execution times include the overhead of transferring data between the host and NPU. All CPU code is compiled at optimisation level three. Experiments conducted in this paper leverage Chess, and all NPU runs are over 16 AIEs (the four columns with a shim tile).

\begin{table}[htbp]
    \begin{center}  
    \begin{tabular}{|c|c|cc|cc|}
    \hline     
       & & \multicolumn{2}{c|}{\textbf{NPU hand written}} & \multicolumn{2}{c|}{\textbf{NPU our approach}} \\
      
      \textbf{Kernel} & \makecell{\textbf{Problem }\\ \textbf{size}} & \makecell{\textbf{\textit{Runtime}} \\ \textbf{\textit{(ms)}}} & \makecell{\textbf{\textit{Lines}} \\ \textbf{\textit{of code}}} & \makecell{\textbf{\textit{Runtime}} \\ \textbf{\textit{(ms)}}} & \makecell{\textbf{\textit{Lines}} \\ \textbf{\textit{of code}}} \\
      \hline
    softmax & 4m & 11.82 & 215 & 11.19 & 24 \\
    relu & 67m & 5.42 & 179 & 5.87 & 5 \\
    saxpy & 67m & 6.12 & 156 & 5.22 & 9 \\
    dot product & 67m & 9.30 & 203 & 9.13 & 9 \\
    l2norm & 67m & 8.77 & 187 & 8.48 & 11 \\
    gemm & 512 & 1.51 & 1540 & 10.56 & 14 \\
    \hline
    \end{tabular}
    \caption{Comparison between hand written kernels from \cite{aie-mlir} using IRON and C++, compared to Fortran and OpenMP using our approach. All using float32 datatype (apart from gemm which uses bf16 as input and float32 as output).}
    \label{tab:iron-cmp}
    \end{center}
    \vspace{-20pt}
\end{table}

We compared foundational kernels important for HPC and AI/ML, and Table \ref{tab:iron-cmp} reports a comparison of kernels using OpenMP following our flow against hand written AIE implementations. The \emph{softmax}, \emph{relu}, \emph{saxpy} and \emph{gemm} hand-written implementations were developed by AMD and from \cite{aie-mlir}, whilst \emph{dot product} and \emph{l2norm} were developed by the authors. It can be seen that performance of our approach is generally comparable to that of hand written AIE implementations, with \emph{gemm} being the outlier because AMD have heavily optimised that kernel but at the cost of code complexity.

\begin{table}[htb]
    \begin{center}  
    \begin{tabular}{|c|c|cc|cc|}
    \hline     
       & & \multicolumn{2}{c|}{\textbf{CPU}} & \multicolumn{2}{c|}{\textbf{NPU our approach}} \\
      
      \textbf{Kernel} & \makecell{\textbf{Problem }\\ \textbf{size}} & \makecell{\textbf{\textit{Runtime}} \\ \textbf{\textit{(ms)}}} & \makecell{\textbf{\textit{Energy}} \\ \textbf{\textit{usage (J)}}} & \makecell{\textbf{\textit{Runtime}} \\ \textbf{\textit{(ms)}}} & \makecell{\textbf{\textit{Energy}} \\ \textbf{\textit{usage (J)}}} \\
      \hline
    softmax & 4m & 10.48 & 0.44 & 11.19 & 0.26 \\
    relu & 67m &  28.27 & 1.04 & 5.87 & 0.17 \\
    saxpy & 67m & 13.79 & 0.39 & 5.22 & 0.15 \\
    dot product & 67m & 9.19 & 0.54 & 9.13 & 0.21 \\
    l2norm & 67m & 5.01 & 0.28 & 8.48 & 0.20 \\
    gemm & 512 & 28.96 & 1.82 & 10.56 & 0.27 \\
    \hline
    \end{tabular}
    \caption{Comparison between all 8 cores of the CPU against using our approach on the NPU. All code OpenMP and float32.}
    \label{tab:kernel-cpu-fp32}
    \end{center}
    \vspace{-20pt}
\end{table}

Lines of code are also reported in Table \ref{tab:iron-cmp}, and it can be seen that the hand written versions require significantly more code to be written than our approach. In calculating this metric we ignore code comments, but includes the IRON graph, C++ compute kernel and C++ host code. This demonstrates one of the major current challenges with writing code for the architecture where to gain performance the programmer must effectively writing three separate pieces of code; the host C++ code using the XRT API, the C++ compute kernel(s) using the AI Engine API, and Python code using IRON for the AIE graph. This is not only time consuming but also requires expertise in all these APIs. By contrast, when using our OpenMP loop based approach the programmer requires just one single codebase and an existing loop that is decorated with OpenMP with the compiler then handling the rest. Consequently, many of the lines of code that are counted under our approach will already exist in the CPU version of the code.

Table \ref{tab:kernel-cpu-fp32} reports a performance and energy usage comparison for these six kernels written in Fortran with OpenMP running multi-threaded on all eight CPU cores and 16 AIEs. It can be observed that performance provided by the NPU is generally competitive against the CPU and regardless energy usage is less when the kernels are run on the NPU. Whilst \emph{gemm} is the outlier as it performs is around 2.8 times faster on the NPU than the eight-core CPU, it should be highlighted that the CPU implementation using Fortran and OpenMP is fairly naive as one would naturally write it and for instance does not leverage known optimisation techniques such as tiling. 

\subsection{Evaluation for scientific computing kernels}
\label{sec:hpc-perf}
We compared two Fortran HPC codes on the CPU against the NPU using our compilation pipeline. Firstly a 2D PW advection scheme from the Met Office's MONC atmospheric model \cite{brown2020highly} used in production runs and optimised for the CPU. The second code is a compute intensive loop from the Shallow Water Equation (SWE) mini-app developed by NCAR and used as a proxy for production workloads. Both kernels perform stencil based computations, an extremely common pattern in scientific computing where calculations are performed for each grid cell and these involve quantities from neighbouring grid cells, a simple example was illustrated in Listing \ref{lst:stencil-ir}.

Table \ref{tab:hpc-hybrid} reports performance (million grid points processed per second, where higher is better) and energy to solution on the AMD Ryzen 8945HS. We leverage a hybrid co-execution strategy where separate chunks of iterations run across the CPU (67\%) and NPU (33\%) concurrently. Float32 is used for all experimental runs, with the addition of the OpenMP \emph{target} pragma being the only code change required for the NPU. This hybrid approach improves performance as the CPU and NPU are processing loop iterations concurrently. Whilst energy to solution is increased compared to running on the NPU alone in Table \ref{tab:kernel-cpu-fp32}, this is still less than the multicore CPU alone.

\begin{table}[htb]
    \begin{center}  
    \begin{tabular}{|c|cc|cc|}
    \hline     
       & \multicolumn{2}{c|}{\textbf{CPU eight cores}} & \multicolumn{2}{c|}{\textbf{Hybrid NPU+CPU}} \\
      
      \textbf{Kernel} & \makecell{\textbf{\textit{Throughput}} \\ \textbf{\textit{(MPts/s)}}} & \makecell{\textbf{\textit{Energy}} \\ \textbf{\textit{usage (J)}}} & \makecell{\textbf{\textit{Throughput}} \\ \textbf{\textit{(MPts/s)}}} & \makecell{\textbf{\textit{Energy}} \\ \textbf{\textit{usage (J)}}} \\
      \hline
    PW advection & 381.84 & 43.11 & 534.06 & 41.43 \\
    SWE & 791.99 & 363.90 & 1008.32 & 315.61 \\    
    \hline
    \end{tabular}
    \caption{Throughout (million grid points per second, higher is better) and energy to solution on CPU and hybrid NPU+CPU using our approach. SWE for 1 million grid points and 4000 iterations, PW advection 268 million grid points. Float32.}
    \label{tab:hpc-hybrid}
    \end{center}
    \vspace{-20pt}
\end{table}

\section{Conclusions}
\label{sec:conclusions}

In this paper we have explored the seamless offloading of general purpose loops, driven by Fortran, to the NPU by decorating them with the OpenMP \emph{target} pragma. By lifting to a tensor representation, we demonstrated that this high-level view of compute provides a rich source of information when targeting the NPU. Driven by guarantees provided by OpenMP, a variety of computing loops can be transformed into a tensor representation, and one is able to then effectively lower through intermediate dialects to AMD's AIE dialects. 

In the main our approach achieves competitive performance to that of hand written codes but at significantly reduced number of lines. The energy to solution is also significantly lower on the NPU compared to either configurations on the CPU. Moreover a hybrid approach, where loops are decomposed across both the CPU and NPU, delivers improved performance whilst still maintaining some energy efficiency benefits.


\section*{Acknowledgments}
This research was supported by an RSE personal research fellowship award number 3271. For the purposes of open access, the author has applied a CC BY public copyright licence to any Author Accepted Manuscript version arising from this submission.
\bibliographystyle{IEEEtran}
\bibliography{references.bib}

\end{document}
\endinput

%% file: setup.tex
\usepackage[frozencache, cachedir=minted-cache]{minted}
\usepackage{etoolbox}

\makeatletter
\ifcsdef{minted@optlistcl@quote}
{
\ifwindows
  \renewcommand{\minted@optlistcl@quote}[2]{%
    \ifstrempty{#2}{\detokenize{#1}}{\detokenize{#1="#2"}}}
\else
  \renewcommand{\minted@optlistcl@quote}[2]{%
    \ifstrempty{#2}{\detokenize{#1}}{\detokenize{#1='#2'}}}
\fi

\newcommand{\minted@def@optcl@novalue}[2]{%
  \define@booleankey{minted@opt@g}{#1}%
    {\minted@addto@optlistcl{\minted@optlistcl@g}{#2}{}%
     \@namedef{minted@opt@g:#1}{true}}
    {\@namedef{minted@opt@g:#1}{false}}
  \define@booleankey{minted@opt@g@i}{#1}%
    {\minted@addto@optlistcl{\minted@optlistcl@g@i}{#2}{}%
     \@namedef{minted@opt@g@i:#1}{true}}
    {\@namedef{minted@opt@g@i:#1}{false}}
  \define@booleankey{minted@opt@lang}{#1}%
    {\minted@addto@optlistcl@lang{minted@optlistcl@lang\minted@lang}{#2}{}%
     \@namedef{minted@opt@lang\minted@lang:#1}{true}}
    {\@namedef{minted@opt@lang\minted@lang:#1}{false}}
  \define@booleankey{minted@opt@lang@i}{#1}%
    {\minted@addto@optlistcl@lang{minted@optlistcl@lang\minted@lang @i}{#2}{}%
     \@namedef{minted@opt@lang\minted@lang @i:#1}{true}}
    {\@namedef{minted@opt@lang\minted@lang @i:#1}{false}}
  \define@booleankey{minted@opt@cmd}{#1}%
      {\minted@addto@optlistcl{\minted@optlistcl@cmd}{#2}{}%
        \@namedef{minted@opt@cmd:#1}{true}}
      {\@namedef{minted@opt@cmd:#1}{false}}
}
\minted@def@optcl@novalue{customlexer}{-x}
}
{
}
\makeatother

\usepackage{tikz}
\usetikzlibrary{arrows}
\usetikzlibrary{shapes}

\tikzset{
  circledstyle/.style={
    shape=circle,
    #1,
    font=\tt\small,
    inner sep=0pt,
    outer sep=0pt,
    minimum size=1.2em,
    text=black,
  }
}

\usepackage[norefs,nocites]{refcheck}

\makeatletter
\newcommand{\refcheckize}[1]{%
  \expandafter\let\csname @@\string#1\endcsname#1%
  \expandafter\DeclareRobustCommand\csname relax\string#1\endcsname[1]{%
    \csname @@\string#1\endcsname{##1}\wrtusdrf{##1}}%
  \expandafter\let\expandafter#1\csname relax\string#1\endcsname
}
\makeatother

\refcheckize{\autoref}

%% file: paper.bbl
\begin{thebibliography}{10}
\providecommand{\url}[1]{#1}
\csname url@samestyle\endcsname
\providecommand{\newblock}{\relax}
\providecommand{\bibinfo}[2]{#2}
\providecommand{\BIBentrySTDinterwordspacing}{\spaceskip=0pt\relax}
\providecommand{\BIBentryALTinterwordstretchfactor}{4}
\providecommand{\BIBentryALTinterwordspacing}{\spaceskip=\fontdimen2\font plus
\BIBentryALTinterwordstretchfactor\fontdimen3\font minus \fontdimen4\font\relax}
\providecommand{\BIBforeignlanguage}[2]{{%
\expandafter\ifx\csname l@#1\endcsname\relax
\typeout{** WARNING: IEEEtran.bst: No hyphenation pattern has been}%
\typeout{** loaded for the language `#1'. Using the pattern for}%
\typeout{** the default language instead.}%
\else
\language=\csname l@#1\endcsname
\fi
#2}}
\providecommand{\BIBdecl}{\relax}
\BIBdecl

\bibitem{brown2023exploring}
N.~Brown, ``Exploring the versal ai engines for accelerating stencil-based atmospheric advection simulation,'' in \emph{Proceedings of the 2023 ACM/SIGDA International Symposium on Field Programmable Gate Arrays}, 2023, pp. 91--97.

\bibitem{klaisoongnoen2024evaluating}
M.~Klaisoongnoen \emph{et~al.}, ``Evaluating versal ai engines for option price discovery in market risk analysis,'' in \emph{Proceedings of the 2024 ACM/SIGDA International Symposium on Field Programmable Gate Arrays}, 2024, pp. 176--182.

\bibitem{hunhoff2025efficiency}
E.~Hunhoff \emph{et~al.}, ``Efficiency, expressivity, and extensibility in a close-to-metal npu programming interface,'' in \emph{2025 IEEE 33rd Annual International Symposium on Field-Programmable Custom Computing Machines (FCCM)}.\hskip 1em plus 0.5em minus 0.4em\relax IEEE, 2025, pp. 85--94.

\bibitem{rodriguez2023fortran}
Rodriguez-Canal \emph{et~al.}, ``Fortran high-level synthesis: Reducing the barriers to accelerating hpc codes on fpgas,'' in \emph{2023 33rd International Conference on Field-Programmable Logic and Applications (FPL)}.\hskip 1em plus 0.5em minus 0.4em\relax IEEE, 2023, pp. 10--18.

\bibitem{aie-ml}
\BIBentryALTinterwordspacing
(2024) Versal adaptive soc aie-ml architecture manual. [Online]. Available: \url{https://docs.amd.com/r/en-US/am020-versal-aie-ml/Overview}
\BIBentrySTDinterwordspacing

\bibitem{bouaziz2025dataflow}
M.~Bouaziz \emph{et~al.}, ``A dataflow overlay for monte carlo multi-asset option pricing on amd versal ai engines,'' in \emph{ISC High Performance 2025 Research Paper Proceedings (40th International Conference)}.\hskip 1em plus 0.5em minus 0.4em\relax Prometeus GmbH, 2025, pp. 1--12.

\bibitem{zhang2023new}
W.~Zhang \emph{et~al.}, ``New filter2d accelerator on the versal platform powered by the ai engine,'' in \emph{International Symposium on Advanced Parallel Processing Technologies}.\hskip 1em plus 0.5em minus 0.4em\relax Springer, 2023, pp. 437--449.

\bibitem{levental2024end}
M.~Levental \emph{et~al.}, ``An end-to-end programming model for ai engine architectures,'' in \emph{Proceedings of the 14th International Symposium on Highly Efficient Accelerators and Reconfigurable Technologies}, 2024, pp. 135--136.

\bibitem{kalkhof2024enabling}
T.~Kalkhof \emph{et~al.}, ``Enabling fpga and ai engine tasks in the hpx programming framework for heterogeneous high-performance computing,'' in \emph{International Symposium on Applied Reconfigurable Computing}.\hskip 1em plus 0.5em minus 0.4em\relax Springer, 2024, pp. 75--89.

\bibitem{brown2025seamless}
N.~Brown \emph{et~al.}, ``Seamless acceleration of fortran intrinsics via amd ai engines,'' in \emph{Proceedings of the 2025 ACM/SIGDA International Symposium on Field Programmable Gate Arrays}, 2025, pp. 185--185.

\bibitem{brown2024fully}
N.~Brown, ``Fully integrating the flang fortran compiler with standard mlir,'' in \emph{SC24-W: Workshops of the International Conference for High Performance Computing, Networking, Storage and Analysis}.\hskip 1em plus 0.5em minus 0.4em\relax IEEE, 2024, pp. 939--949.

\bibitem{huthmann2020openmp}
J.~Huthmann \emph{et~al.}, ``{OpenMP device offloading to FPGAs using the Nymble infrastructure},'' in \emph{International Workshop on OpenMP}.\hskip 1em plus 0.5em minus 0.4em\relax Springer, 2020, pp. 265--279.

\bibitem{openmpfpga}
G.~Rodriguez-Canal \emph{et~al.}, ``An mlir pipeline for offloading fortran to fpgas via openmp,'' 2025.

\bibitem{brauckmann2025tensorize}
A.~Brauckmann \emph{et~al.}, ``Tensorize: Fast synthesis of tensor programs from legacy code using symbolic tracing, sketching and solving,'' in \emph{Proceedings of the 23rd ACM/IEEE International Symposium on Code Generation and Optimization}, 2025, pp. 15--30.

\bibitem{sc25-intrinsic}
N.~Brown and G.~Rodriguez-Canal, ``Programmer productivity and performance on amd’s ai engines: Offloading fortran intrinsics via mlir a case-study,'' 2025.

\bibitem{tosa}
\BIBentryALTinterwordspacing
(2025) Tensor operator set architecture (tosa). [Online]. Available: \url{https://www.mlplatform.org/tosa/tosa_spec.html}
\BIBentrySTDinterwordspacing

\bibitem{aie-vect}
\BIBentryALTinterwordspacing
(2025) Aie automatic vectorization. [Online]. Available: \url{https://github.com/Xilinx/mlir-aie/blob/main/docs/AIEVectorization.md}
\BIBentrySTDinterwordspacing

\bibitem{aie-mlir}
\BIBentryALTinterwordspacing
(2025) Iron api and mlir-based ai engine toolchain. [Online]. Available: \url{https://github.com/Xilinx/mlir-aie}
\BIBentrySTDinterwordspacing

\bibitem{brown2020highly}
N.~Brown \emph{et~al.}, ``A highly scalable met office nerc cloud model,'' \emph{arXiv preprint arXiv:2009.12849}, 2020.

\end{thebibliography}
